\documentclass[english,showpacs,preprintnumbers,amsmath,amssymb,prb,floatfix]{revtex4}
\usepackage[T1]{fontenc}
\usepackage[latin9]{inputenc}
\usepackage{array}
\usepackage{rotating}
\usepackage{multirow}
\usepackage{amsmath}
\usepackage{amssymb}
\usepackage{graphicx}
\usepackage{esint}

\makeatletter

\providecommand{\tabularnewline}{\\}

\@ifundefined{textcolor}{}
{%
 \definecolor{BLACK}{gray}{0}
 \definecolor{WHITE}{gray}{1}
 \definecolor{RED}{rgb}{1,0,0}
 \definecolor{GREEN}{rgb}{0,1,0}
 \definecolor{BLUE}{rgb}{0,0,1}
 \definecolor{CYAN}{cmyk}{1,0,0,0}
 \definecolor{MAGENTA}{cmyk}{0,1,0,0}
 \definecolor{YELLOW}{cmyk}{0,0,1,0}
}

\makeatother

\usepackage{babel}
\begin{document}

\title{Mode engineering with a one-dimensional superconducting metamaterial}

\author{Masahiko Taguchi,$^{1,2}$ Denis M. Basko,$^{2}$ and Frank W. J.
Hekking$^{2}$}

\affiliation{$^{1}$Graduate School of Pure and Applied Sciences, University of
Tsukuba, Ibaraki 305-8571, Japan}

\affiliation{$^{2}$LPMMC, CNRS/University Joseph Fourier, BP 166, 38042 Grenoble,
France}

\date{\today
}
\begin{abstract}
We propose a way to control the Josephson energy of a single Josephson
junction embedded in one-dimensional superconducting metamaterial: an inhomogeneous superconducting loop, made out of a superconducting nanowire or a chain of Josephson junctions.
The Josephson energy is renormalized by the electromagnetic modes propagating along the loop. We study
the behaviour of the modes as well as of their frequency spectrum when the capacitance and the inductance along the loop are spatially modulated. We show that, depending on the amplitude of the modulation, the
renormalized Josephson energy is either larger or smaller than the one found for a homogeneous loop. Using typical experimental parameters for Josepshon junction chains and superconducting nanowires, we conclude that this mode-engineering can be achieved with currently available metamaterials.
\end{abstract}


\maketitle

\section{Introduction}

It is well-known that low-dimensional superconductors such as Josephson junctions~\cite{Dahm68}, thin films~\cite{Buisson94}, and narrow wires~\cite{Camarota01} sustain plasma excitations: low-frequency collective oscillations of the superfluid charge density. This is possible as the low dimensionality renders Coulomb interactions less effective, thereby reducing the characteristic plasma frequency to values well below the superconducting gap. As a result, at low temperatures, the damping due to quasi-particles is negligible. In extended homogeneous systems, these oscillations acquire a propagating character with a dispersion relation that is linear for one-dimensional systems and of the square-root type for two dimensional systems~\cite{Mooij85}.

Propagating plasma modes have been observed in a variety of systems, including wire networks~\cite{Parage98}, Josephson junction arrays~\cite{Masluk12,Weissl15}, and high-$T_c$ superconductors~\cite{Fertig90,Dunmore95}. They play a fundamental role in the collective behaviour of low-dimensional superconductors. For example, propagating plasma modes provide the quantum fluctuations responsible for the occurrence of quantum phase-slips that eventually drive the superfluid-insulator transition~\cite{Arutyunov08,Fazio01,Matveev02,Rastelli13}. This is relevant in view of recent experiments probing quantum phase-slips in nanowires~\cite{Astafiev12,Peltonen13} and in Josephson junction chains~\cite{Pop10,Manucharyan12}. The propagating plasma modes are also at the origin of the damping of vortex excitations in Josephson junction arrays~\cite{Fazio01}.

Using modern fabrication techniques, the parameters characterizing the properties of the collective plasma excitations in low-dimensional superconductors can be chosen from a relatively wide range. This is especially true for nanostructured superconducting circuits such as Josephson junction chains and arrays. SQUID-based chains for instance sustain propagating plasma modes with a phonon-like dispersion, whose group velocity can be tuned {\em in situ} with the help of the applied external flux~\cite{Weissl15}. This fact has already been exploited successfully in various experiments where Josephson junction chains are used as metamaterials providing dedicated electromagnetic environments. Examples are the use of Josephson junction chains to obtain a tunable environment in order to study the Bloch band dynamics of a single Josephson junction~\cite{Corlevi06, Ergul13,Weissl15} and to observe the dynamical Casimir effect~\cite{Lahteenmaki13}.

In this paper, we show how the controlled introduction of inhomogeneities in a superconducting metamaterial can enhance its use as a dedicated environment. In general, a spatial modulation of the metamaterial's parameters will affect the propagating modes as well as their frequency spectrum. An example is the frequency shift found for the modes of a disordered Josephson junction chain, together with the formation of localized modes~\cite{Basko13}. This offers the possibility of mode engineering: positioning the mode frequencies and controlling the mode amplitudes locally along the metamaterial in order to optimize its frequency response as well as the way it couples to the system one wishes to study.

Specifically, we consider a one-dimensional superconducting metamaterial, forming a loop closed by a small Josephson junction. The loop is threaded by a magnetic flux which induces a persistent current in the loop, see Fig.~\ref{fig:configuration}. The propagating plasma modes along the loop give rise to quantum fluctuations of the phase difference across the  small junction, thereby renormalizing the junction's Josephson energy, and hence the persistent current response~\cite{Hekking97}. Comparing a homogeneous loop with a spatially, periodically modulated one, we show that the renormalization can be made either significantly weaker or significantly stronger, depending on the sign of the modulation amplitude, see, {\em e.g.}, Fig.~\ref{fig:renjoscharge}. We find that this effect is mainly due to the effect of the modulation on the behaviour of modes close to the junction.

\section{The model\label{sec:The-model}}

\begin{figure}[h]
\includegraphics[scale=0.4,angle=0]{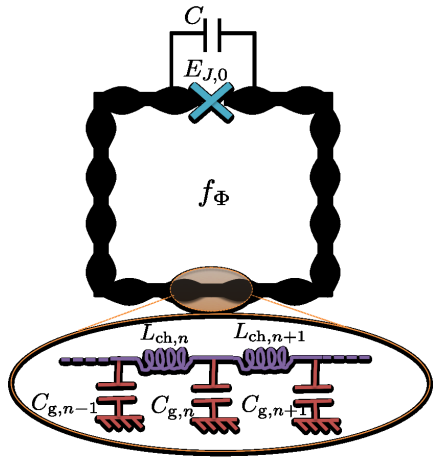}
\caption{(Color online) Single Josephson junction with Josephson energy $E_{J,0}$ and capacitance $C$, embedded in a loop of length $L$ made out of one-dimensional superconducting metamaterial. The metamaterial can be either a thin superconducting wire whose parameters (such as cross-sectional area or distance to a nearby screening gate) are spatially modulated, or a chain of Josephson junctions (see inset) with spatially distributed capacitances $C_{g,n}$ and inductances $L_{ch,n}$.\label{fig:configuration}}
\end{figure}

We consider the system depicted in Fig.~\ref{fig:configuration}. It consists of a one-dimensional superconducting metamaterial of length $L$, closed by a small Josephson junction to form a loop. The metamaterial can be characterized by a space-dependent capacitance per unit length $c(x)$ with respect to ground as well as by a space-dependent kinetic inductance per unit length $l(x)$. The junction has a bare Josephson coupling energy $E_{J,0}$ and charging energy $E_C = (2 e)^2/2C$, where $C$ is the junction capacitance. The loop is threaded by a magnetic flux $\Phi$. For later use we define the reduced flux $f_\Phi = 2 \pi \Phi/\Phi_0$, where $\Phi_0 = h/2e$ is the superconducting flux quantum.

\subsection{Hamiltonian}
\label{sec:Hamiltonian}

The low-energy properties of the system can be described
in terms of the superconducting phase along the loop, $\phi (x)$, and its conjugate momentum, the space-dependent superfluid density $\Pi(x)$, such that $[\phi(x),\Pi(x')] = i \delta(x-x')$.  The Hamiltonian $H$ describing the system then reads
\begin{equation}
 H =  \int_{0}^{L}dx \left\{e_c(x) \Pi^2(x) + e_l(x) \left[\frac{\partial \phi(x)}{\partial x} - \frac{f_\Phi}{L}\right]^2 \right\} +\frac{\left[e_c(L)\Pi(L)-e_c(0)\Pi(0)\right]^{2}}{E_{C}} -E_{J,0}\cos\left[\phi\left(L\right)-\phi\left(0\right)\right].\label{eq:hamiltonian}
\end{equation}
The first term on the right-hand side describes the one-dimensional, inhomogeneous superconducting metamaterial; it is a sum of two contributions. The first one, quadratic in the superfluid density, corresponds to the electrostatic energy, where $e_c(x) = (2 e)^2/2c(x)$. The second one, quadratic in the difference of the phase gradient and the magnetic flux, corresponds to the inductive energy associated with the supercurrent in the metamaterial. Here we define $e_l(x) = (\Phi_0/2\pi)^2/2 l(x)$. Throughout this paper we will assume the phase fluctuations in the metamaterial to be small, so that nonlinear phenomena such as phase-slips can be ignored. This is achieved by imposing the condition $e_c <e_l$. For later use, we also define the space-dependent plasma velocity $v_\mathrm{pl}(x) = \sqrt{1/l(x) c(x)}$ and the zero-frequency dimensionless conductance
$g(x) = \sqrt{c(x)/l(x)} \pi \hbar/(2 e)^2$ of the inhomogeneous metamaterial. The condition $e_c <e_l$ then translates into the condition $g>1$. The quadratic Hamiltonian describing the metamaterial constitutes a low-energy description, valid for energies smaller than some cut-off energy $E_\mathrm{max}$. The resulting long-wavelength theory describes spatial variations of phase and density along the metamaterial on length scales larger than the cut-off length $L_0 = \hbar v_\mathrm{pl}/E_\mathrm{max}$.

The remaining terms in Eq.~(\ref{eq:hamiltonian}) describe the charging energy and the Josephson energy
of the single Josephson junction, respectively. For simplicity, we will neglect
the junctions's capacitance $C$ throughout this paper. This is possible as long as the capacitance of the metamaterial close to the junction is larger than $C$. The smallest part of the metamaterial involved in the plasma oscillation is given by the cut-off length $L_0 = \hbar v_\mathrm{pl}/E_\mathrm{max}$. Then we can set $C=0$ if $L_0 c \gg C$ or, in other words, when $g E_C \gg E_\mathrm{max}$.

The parameters $E_\mathrm{max}$, $c$, and $l$ can be related to the parameters describing the actual system realizing the metamaterial. If it is made out of a narrow superconducting wire with cross-sectional area $S = w^2$ embedded in a dielectric material with dielectric constant $\epsilon$ and sitting at a distance $d$ from a screening gate, $1/c = 2 \ln(d/w)/\epsilon$ and $1/l = 2 n_s e^2 S/m$,  with $e$ the electron charge, $m$ the electron mass, and $n_s$ the density of the superconducting condensate. The quantities $c$ and $l$ become space-dependent if one modulates the cross-sectional area, $S(x)$, and the distance to the screening gate, $d(x)$. The cut-off energy $E_\mathrm{max}$ is given by the superconducting gap $\Delta$. The metamaterial can be considered one-dimensional as long as the cross-sectional dimension $w$ is smaller than the penetration depth, $w < \lambda_L$, where $\lambda_L = \sqrt{mc_\mathrm{light}^2/(4 \pi n_s e^2)}$ with $c_\mathrm{light}$ the speed of light.

The metamaterial can also be realized using a Josephson junction chain consisting of junctions with Josephson inductance $L_{ch} = (\Phi_0/2 \pi)^2/E_{J,ch}$, where $E_{J,ch}$ is the Josephson energy, connecting islands of linear dimension $a$ and with ground capacitance $C_{g}$. Then we have  $c = C_{g}/a$ and $l=L_{ch}/a$. The condition $g > 1$ assures that the characteristic frequency $\omega_0 = \sqrt{1/L_{ch} C_g} < E_{J,ch}/\hbar$, such that nonlinearities associated with the Josephson effect can be ignored. The parameters describing the chain become space-dependent by modulating the size of the junctions and the islands along the chain, such that $C_g \to C_{g,n}$ and $L_{ch} \to L_{ch,n}$, see Fig.~\ref{fig:configuration}. We assume the chain's modes to have frequencies well below the superconducting gap $\Delta$ of the islands. We also ignored the capacitance $C_{ch}$ of the junctions forming the chain. This implies that we assume the mode frequencies to be smaller than the Josephson plasma frequency $\hbar \omega_{p,ch} = \sqrt{8 E_{J,ch} E_{C,ch}}$. The cut-off energy $E_\mathrm{max}$ is therefore determined by the smallest of these two energies $\Delta$ and $\hbar \omega_{p,ch}$.

We summarize the parameter identification for Josephson junction chains and superconducting nanowires in table~\ref{table}.

\begin{table}[h]
\begin{tabular}{|c|c|c|}
\hline
 & JJ-chains & superconducting nanowire\tabularnewline
\hline
\hline
inverse capacitance per unit length & $a/C_{g}$ & $2 \ln(d/w)/\epsilon$\tabularnewline
\hline
inductance per unit length & $L_{ch}/a$ & $m/(2 n_s e^2 S)$\tabularnewline
\hline
plasma velocity $v_{\mathrm{pl}}$ & $a/\sqrt{L_{ch}C_{g}}$ & $c_\mathrm{light}(w/\lambda_L)\sqrt{ \ln (d/w)/\pi \epsilon}$\tabularnewline
\hline
dimensionless conductance $g$ & $\sqrt{C_{g}/L_{ch}}\pi \hbar/(2 e)^2$  & $ (w/8 \lambda_L)(\hbar c_\mathrm{light}/e^2) \sqrt{\pi \epsilon/ \ln(d/w)}$\tabularnewline
\hline
\end{tabular}

\caption{Metamaterial parameters in terms of the parameters of JJ-chains and superconducting nanowires. \label{table}}
\end{table}

\subsection{Classical phase configuration}
\label{sec:classical}

To gain some insight in the behaviour of the system, we start by determining the classical phase configuration $\phi_\mathrm{cl}(x)$ along the loop, {\em i.e.}, the configuration for which the sum of the potential energy terms of Hamiltonian (\ref{eq:hamiltonian}) is stationary. It satisfies the differential equation
\begin{equation}
\frac{d}{dx}\left[ e_l(x) \left(\frac{d \phi_\mathrm{cl}(x)}{dx} - \frac{f_\Phi}{L} \right)\right] = 0 \label{eq:class_conf},
\end{equation}
implying that $e_l(x) [d \phi_\mathrm{cl}(x)/dx - f_\Phi/L] = A $ where $A$ is a constant. Thus, the classical phase configuration is given by
\begin{equation}
\phi_\mathrm{cl}(x) = \phi_\mathrm{cl}(0) + A \int _0^x \frac{dx'}{e_l(x')} + f_\Phi x/L \label{eq:phi_class}.
\end{equation}
The constant $A$ is determined by minimizing the total potential energy. Introducing $\phi_0 \equiv \phi_\mathrm{cl}(L) - \phi_\mathrm{cl}(0) = f_\Phi + A/E_l$, where we define the total inverse inductive energy of the metamaterial
\begin{equation}
\frac{1}{E_l} = \int \limits _0 ^L \frac{dx}{e_l(x)} \label{eq:ind_energy},
\end{equation}
the minimization condition reads
\begin{equation}
\frac{L^*}{L}(\phi_0 - f_\Phi) + \sin \phi_0 =0.
\label{eq:min_pot_en}
\end{equation}
Here
$L^{*}/L = 2 E_l/E_{J,0}$; in other words, $L^*$ is the length for which the energy of the supercurrents in the metamaterial and the
Josephson energy of the junction are of the same order.

\begin{figure}[h]
\includegraphics[scale=0.6,angle=0]{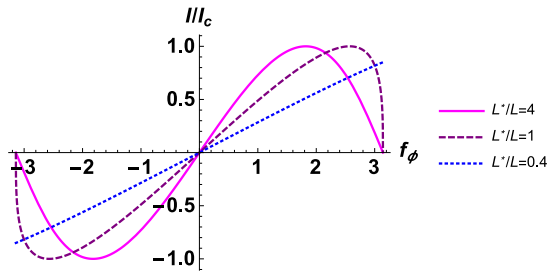}
\caption{(Color online) Persistent current $I/I_c$ as a function of flux $f_\phi$ for a classical loop. Curves from top to bottom correspond to $L^*/L =$ 4 (solid pink), 1 (dashed purple), and 0.4 (dotted blue), respectively.\label{fig:cur_flux}}
\end{figure}

Calculating the total potential energy $U$ for the configuration (\ref{eq:phi_class}) using the condition (\ref{eq:min_pot_en}), one finds that it depends periodically on the magnetic flux $f_\Phi$. It is straightforward to calculate the persistent current $I(f_\Phi) = (2e/\hbar) dU/d f_\Phi$ induced in the loop by the external flux. It is given by $I(f_\Phi) = I_c \sin \phi_0(f_\Phi)$, where $I_c = 2 e E_{J,0}/\hbar$ and $\phi_0(f_\Phi)$ is found by solving Eq.~(\ref{eq:min_pot_en}). In Fig.~\ref{fig:cur_flux} we plot the expected current - flux relationship for various values of the ratio $L^*/L$. For short loops $L^*/L \gg 1$ the response is purely sinusoidal. Indeed, according to (\ref{eq:min_pot_en}), $\phi_0 \approx f_\Phi$ in this limit and the phase-difference across the junction is completely determined by the flux threading the loop. In the opposite limit $L^*/L \ll 1$, the persistent current shows a saw-tooth like behaviour as a function of $f_\Phi$. In this limit, the Josephson junction pins the phase difference $\phi_0$ to values close to integer multiples of $2 \pi$. When $f_\Phi$ is increased from $-\pi$ to $\pi$, the phase difference $\phi_0 \approx 0$ will increase slowly, $\phi_0 = L^* f_\Phi/L$, causing a linear dependence on $f_\Phi$, $I(f_\Phi) = L^* I_c f_\Phi/L $. When $f_\Phi$ reaches the value $\pi$, $\phi_0$ jumps from the value $L^* \pi/L$ to the value $2 \pi - \pi L^*/L$, causing the persistent current to jump from $\pi L^* I_c/L$ to $-\pi L^*I_c/L$. This jump is again followed by a linear increase $I(f_\Phi) = L^* I_c/L(f_\Phi - 2 \pi)$. A similar jump occurs for $f_\Phi = - \pi$.

\subsection{Quantum fluctuations}
\label{sec:quantum}

We will now include the effect of the electrostatic energy stored in the metamaterial. This energy causes the phase to fluctuate around its classical value $\phi_\mathrm{cl}$. Setting $\phi(x) = \phi_\mathrm{cl}(x) + \chi(x)$, we obtain the effective Hamiltonian $H_\mathrm{eff} = H_ 0 + H_J$ that governs the behaviour of the quantum fluctuations $\chi(x)$, where
\begin{eqnarray}
 H_0 & = & \int_{0}^{L}dx \left[e_c(x) \Pi^2(x) + e_l(x) \left(\frac{\partial \chi(x)}{\partial x}\right)^2 \right], \label{eq:H0} \\
 H_J&=& - E_{J,0}\left\{ \left[\chi(L) - \chi(0)\right] \sin \phi_0 + \cos\left[\phi_0 + \chi\left(L\right)-\chi\left(0\right)\right]\right\}.
\label{eq:HJ}
\end{eqnarray}

In the remainder of this paper, we will treat the Josephson junction as a perturbation, assuming the Josephson coupling energy $E_{J,0}$ to be small. This means that we focus on relatively short loops, such that $L< L^*$. We diagonalize the unperturbed effective Hamiltonian $H_0$ with the help of the mode expansions
\begin{eqnarray}
\chi\left(x\right) & = & \sqrt{\frac{1}{2}} \sum_{n=0}^{n_\mathrm{max}}\sqrt{\frac{2 e_{c}\left(x\right)}{\hbar \omega_n}}\psi_{n}\left(x\right)\left[a_{n}^{\dagger}+a_{n}\right],\label{eq:quantum fluctuation}\\
\Pi\left(x\right) & = & i\sqrt{\frac{1}{2}}\sum_{n=0}^{n_\mathrm{max}}\sqrt{\frac{\hbar \omega_n}{2 e_{c}\left(x\right)}}\psi_{n}\left(x\right)\left[a_{n}^{\dagger}-a_{n}\right],\label{eq:conjugate mometum}
\end{eqnarray}
where the cut-off value $n_\mathrm{max}$ is defined through $\hbar \omega_{n_{\mathrm{max}}} = E_\mathrm{max}$.
This yields the correct commutation relation $\left[\chi\left(x\right),\Pi\left(x'\right)\right]=i \delta\left(x-x'\right)$
when imposing the completeness of $\psi_{n}\left(x\right)$,
\begin{equation}
\sum_n \psi_{n}\left(x\right)\psi_{n}\left(x'\right)=\delta\left(x-x'\right)\label{eq:completeness}.
\end{equation}
The unperturbed Hamiltonian $H_0$ takes its diagonal form
\begin{equation}
H_0 = \sum_{n=0}^{n_\mathrm{max}}\hbar \omega_n \left(a^\dagger_n a_n + 1/2 \right),\label{eq:diagH_0}
\end{equation}
provided the functions $\psi_n\left(x\right)$
satisfy the inhomogeneous wave equation
\begin{equation}
\sqrt{e_{c}\left(x\right)}\frac{d}{dx}\left\{e_l\left(x\right)\frac{d}{dx}\left[ \sqrt{e_{c}\left(x\right)}\psi_{n}\left(x\right)\right] \right\}=-\hbar^2 \omega_{n}^{2}\psi_{n}\left(x\right)/4\label{eq:condition for inhomogecity},
\end{equation}
together with the orthogonality relation
\begin{equation}
\int_{0}^{L}\psi_{n}\left(x\right)\psi_{m}\left(x\right) dx= \delta_{n,m}.\label{eq:orthogonality condition}
\end{equation}
Finally, since no current flows through the ends of the metamaterial in the absence of the junction, we impose the boundary condition
\begin{equation}
\left.\frac{d\psi_{n}\left(x\right)}{dx}\right|_{x=0,L}=0\label{eq:boundary condition}.
\end{equation}
For later use, it is convenient to introduce the auxiliary functions $\Psi_n(x) = \sqrt{e_c(x)} \psi_n(x)$. They satisfy the inhomogeneous wave equation
\begin{equation}
e_{c}\left(x\right)\frac{d}{dx}\left[e_l\left(x\right)\frac{d \Psi_n(x)}{dx} \right]=-\hbar^2 \omega_{n}^{2}\Psi_{n}\left(x\right)/4\label{eq:Psi_n}.
\end{equation}
The functions $\Psi_{n}\left(x\right)$ are subjected to the weighted orthogonality condition
\begin{equation}
\int_{0}^{L} [e_c(x)]^{-1} \Psi_{n}\left(x\right) \Psi_{m}\left(x\right) dx=\delta_{n,m}.\label{eq:[c]orthogonality condition Psi}
\end{equation}
We can express the fluctuating part of the phase $\chi(x)$ directly in terms of the functions $\Psi_n$,
\begin{equation}
\chi\left(x\right)  =  \sum_{n=0}^{n_\mathrm{max}}\sqrt{\frac{1}{\hbar \omega_n}}\Psi_{n}\left(x\right)\left[a_{n}^{\dagger}+a_{n}\right].\label{eq:quantum fluctuation_Psi}\\
\end{equation}

\subsection{Effect of quantum fluctuations: suppression of the persistent current}
In this paper, we will analyze the effect of the quantum fluctuations on the persistent current response of the loop in the situation where the Josephson junction can be treated as a perturbation. This is achieved by computing the average of Josephson part $H_J$, Eq.~(\ref{eq:HJ}), of the effective Hamiltonian $H_\mathrm{eff}$ with respect to the fluctuations $\chi$, Eq.~(\ref{eq:quantum fluctuation_Psi}), governed by the unperturbed Hamiltonian $H_0$, Eq.~(\ref{eq:diagH_0}). The average of the first term on the right hand side of Eq.~(\ref{eq:HJ}) vanishes. The average of the second term implies computing $\langle \cos[\phi_0 + \chi(L) - \chi(0)]\rangle = \cos \phi_0 \langle \cos[\chi(L) - \chi(0)]\rangle$. Therefore, $\langle H_J\rangle = -E_J \cos \phi_0$, where $E_J = E_{J,0} \langle \cos[\chi(L) - \chi(0)]\rangle$. Using the identity $\langle \exp\{i{[\chi(L) - \chi(0)]}\}\rangle = \exp\{-\langle[\chi(L) - \chi(0)]^2\rangle/2\}$, we obtain $E_J = E_{J,0} \exp(-\sum_{n=0}^{n_\mathrm{max}} \lambda_n^2/2)$ with
\begin{equation}
\lambda_n = \sqrt{\frac{1}{\hbar \omega_n}}[\Psi_{n}\left(L\right)- \Psi_{n}\left(0\right)] \sqrt{\coth(\beta \hbar \omega_n/2)}.\label{eq:lambda_n}\\
\end{equation}
Here we used the fact that $\langle a^\dagger_n a_n\rangle = [\exp(\beta \hbar \omega_n) -1]^{-1}$ with $\beta = 1/k_B T$. This implies that the persistent current response of the loop will be qualitatively similar to but quantitatively different from the one found above in the classical limit: the bare Josephson coupling $E_{J,0}$ energy should be replaced by its renormalized value $E_J$.

\begin{figure}[h]
\includegraphics[scale=0.6,angle=0]{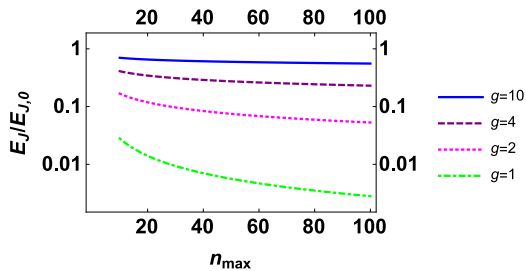}
\caption{(Color online) Ratio $E_J/E_{J,0}$, obtained from Eq.~(\ref{eq:homrenres}), as a function of $n_\mathrm{max}$ for a Josephson junction embedded in a homogeneous metamaterial. Curves from bottom to top correspond to $g=1$ (dashed-dotted green), 2 (dotted pink), 4 (dashed purple) and 10 (solid blue). \label{fig:renormej}}
\end{figure}

The renormalized Josephson energy $E_J$ has been computed in Ref.~[\onlinecite{Hekking97}] for a homogeneous metamaterial. Setting $c(x) = c_0$ and $l(x) = l_0$, Eq.~(\ref{eq:Psi_n}) reduces to the homogeneous wave equation
\begin{equation}
v_\mathrm{pl}^2 \frac{d^2 \Psi_n(x)}{dx ^2} = - \omega_{n}^{2}\Psi_{n}\left(x\right),\label{eq:hom_case}
\end{equation}
where $v_\mathrm{pl}^2  = 1/l_0 c_0$. Its solutions are
\begin{equation}
\Psi_n(x) = \sqrt{2 e_c/L} \cos (k_n x), \label{eq:hom_wave}
\end{equation}
with $e_c = (2 e)^2/2 c_0$, $k_n = n\pi/L$ and $\omega_n = v_\mathrm{pl} k_n$. This solution satisfies the weighted orthogonality condition (\ref{eq:[c]orthogonality condition Psi}) and is compatible with the boundary condition (\ref{eq:boundary condition}). Taking the zero-temperature limit, $T=0$, we find
\begin{equation}
\lambda_n = \sqrt{\frac{1}{g n}} [(-1)^n - 1] ,\label{eq:lambda_n_hom}\\
\end{equation}
where $g = \sqrt{c_0/l_0} \pi \hbar/4e^2$. Using the results (\ref{eq:[c]replacement1}) and (\ref{eq:[c]replacement2}) from Appendix \ref{sec:sums}, we find
\begin{equation}
\sum \limits_{n=0}^{n_\mathrm{max}} \lambda_n^2 = (2/g)[\ln n_{\mathrm{max}}+\gamma + \ln 2 +\mathcal{O}(1/n_{\mathrm{max}})],
\end{equation}
where $\gamma = 0.5772 \ldots$ is Euler's constant.
As a result
\begin{equation}
E_J = E_{J,0} (1/2 e^\gamma n_\mathrm{max})^{1/g}.
\label{eq:homrenres}
\end{equation}
We plot the ratio $E_J/E_{J,0}$ as a function of $n_\mathrm{max}$ for various values of $g$ in Fig.~\ref{fig:renormej}.
Below we will see how this result is modified for an inhomogeneous metamaterial. We will see in particular that, depending on the modulation, the renormalization of $E_{J,0}$ down to $E_J$ can be either weaker or stronger than the one found for a homogeneous metamaterial.

\section{Modulating the capacitance\label{sec:Modulating capacitance}}

In this section, we consider a metamaterial for which the capacitance $c(x)$ is modulated along the loop, yielding a modulation of the charging energy $e_c(x)$. Setting $e_l(x) = e_l = (\Phi_0/2 \pi)^2/2 l_0$ constant, Eq.~(\ref{eq:Psi_n}) takes the form
\begin{equation}
\frac{d^{2}\Psi_{n}\left(x\right)}{dx^{2}}+\frac{\omega_{n}^{2}}{v_\mathrm{pl}^2(x)}\Psi_{n}\left(x\right)=0,\label{eq:[c]equation1}
\end{equation}
where we used the fact that $\sqrt{e_l(x) e_c(x)} = \hbar v_\mathrm{pl}(x)/2$. For simplicity, we restrict ourselves to modulations such that
\begin{equation}
\left.\frac{de_{c}\left(x\right)}{dx}\right|_{x=0,L}=0.\label{eq:[c]boundary condition}
\end{equation}
Then the boundary conditions for $\Psi_{n}\left(x\right)$
are $d\Psi_n\left(x\right)/d x|_{x=0,L}=0$.

To be specific, we consider a periodic modulation of the capacitance, such that
\begin{equation}
c\left(x\right)=c_0 (1-t\cos k_{m}x),\label{eq:modulating capacitance}
\end{equation}
where $c_0$ is the average capacitance per unit length of the metamaterial, $k_{m}=m\pi/L$ with $m$ an integer, and $t$
is the relative modulation amplitude such that $|t| <1$. The modulation period is thus given by $2 L/m$. It is easy to verify that (\ref{eq:modulating capacitance}) yields a modulation of $e_c(x)$ that is compatible with Eq.~(\ref{eq:[c]boundary condition}).

With the modulation
Eq.~(\ref{eq:modulating capacitance}), Eq.~(\ref{eq:[c]equation1}) becomes
\begin{equation}
\frac{d^{2}\Psi_{n}\left(x\right)}{dx^{2}}+E_{n}\left[1-t\cos (k_{m}x)\right]\Psi_{n}\left(x\right)=0,\label{eq:[c]equation2}
\end{equation}
where we defined $E_{n}=\left(\omega_{n}/v_\mathrm{pl}\right)^{2}$ with $v^2_\mathrm{pl} = 1/l_0 c_0$. Equation (\ref{eq:[c]equation2}) is similar to the Schr\"odinger equation. However, here the
eigenvalue $E_n$ multiplies also the potential $t\cos(k_m x)$. For this
reason, even for small $t$, the potential term can become comparable
to the kinetic term for sufficiently high energies. For $t=0$, we obtain the
homogeneous wave equation (\ref{eq:hom_case}) discussed above.

\subsection{Perturbation theory with respect to $t$}
\label{subsec:perturbation}

We wish to determine the eigenfunctions $\Psi_n(x)$ and the eigenvalues $E_n$ of Eq.~(\ref{eq:[c]equation2}). Since the modulation term is periodic in $x$, we expect the spectrum to consist of bands and gaps. As long as the relative amplitude $t$ is small, $t \ll 1$, we can use standard perturbative methods to obtain $E_n$ and $\Psi_n$. Since the perturbation mixes the unperturbed mode $\cos (k_n x)$ with the modes $\cos (k_{n-m} x)$ and $\cos (k_{n+m} x)$, its dominant feature is to open a gap $\propto |t|$ in the spectrum $E_n$ at the degeneracy point  $k_n = \pi m/2 L = k_m/2$. Other gaps exist, but they scale with higher powers of $t$ and hence will be ignored in the following. We therefore set
\begin{equation}
\Psi_{n}\left(x\right)=A_{n}\cos (k_{n}x)+B_{n}\cos (k_{n-m}x) + C_n \cos (k_{n+m}x),\label{eq:[c]three cosine}
\end{equation}
and substituting this into Eq.~(\ref{eq:[c]equation2}), we find the set of equations
\begin{eqnarray}
A_{n}\left(k_{n}^{2}-E_{n}\right)+\frac{tE_{n}}{2}(B_{n}+C_n) & = & 0,\label{eq:[c]condition 1}\\
\frac{tE_{n}}{2}A_{n}+\left(k_{n-m}^{2}-E_{n}\right)B_{n} & = & 0,\label{eq:[c]condition 2}\\
\frac{tE_{n}}{2}A_{n}+\left(k_{n+m}^{2}-E_{n}\right)C_{n} & = & 0.\label{eq:[c]condition 3}
\end{eqnarray}
From Eq.~(\ref{eq:[c]condition 3}) we see that
\begin{equation}
C_{n} = \frac{tE_{n}/2}{E_{n}-k_{n+m}^{2}}A_{n}.\label{eq:[c]coefficient2}
\end{equation}
The denominator of this expression is never small, as both $n,m$ are
positive, so $C_n=\mathcal{O}(t)$. Its substitution into Eq.~(29) produces a
correction to $E_n$ which is $\mathcal{O}(t^2)$, so we neglect it in the
following.
At the same time, the difference $k_{n-m}^2-E_n$ in Eq.(30) may
become small at $n\approx m/2$, so degenerate perturbation theory for
$A_n$ and $B_n$ should be used, which gives the characteristic equation
for $E_n$
\begin{equation}
(k_n^2 -E_n)(k_{n-m}^2- E_n) - t^2 E_n^2/4 = 0.
\end{equation}
This implies that
\begin{equation}
E_{n,\pm} = \frac{(k_n^2 + k_{n-m}^2) \pm \sqrt{(k_n^2 -k_{n-m}^2)^2 + t^2 k_n^2 k_{n-m}^2}}{2},
\label{eq:eigenvalues}
\end{equation}
where we ignored corrections of order $t^2$ (note that the square root
is of the order $t$ if the difference $k_n^2-k_{n-m}^2$ is small). The spectrum
is given by $E_{n,-}$ for $n<m/2$ and by $E_{n,+}$ for $n>m/2$. For $n=m/2$,
we find the expected gap $|t|m^2\pi^2/(4L^2)$. The gap region extends over
a range $\delta n\sim mt$ around $n=m/2$.

Equation (\ref{eq:[c]coefficient2}) works well in the gap region, however, it has problems at
large~$n$. Indeed, while the numerator $\sim n^2t$, the denominator
$\sim mn$. Thus, for $n>m/t$, Eq.~(\ref{eq:[c]coefficient2}) no longer represents a small correction,
so the perturbation theory breaks down. This breakdown is a consequence of
the feature of Eq.~(\ref{eq:[c]equation2}), discussed above:  the weak modulation is multiplied
by the eigenvalue $E_n$, so it does not represent a small perturbation when
$E_n\propto{n}^2$ is large. Thus, modes with large $n$ should be treated
differently, which will be done in the next subsection. For the moment, we
restrict our consideration to the modes with $n\ll m/t$, and proceed with the
calculation of $E_J(t)$.

We next determine the coefficients $A_n$, $B_n$, and $C_n$. In the vicinity of the degeneracy point such that $n$ is within a distance $\delta n$ of $m/2$, the modulation mixes predominantly the modes $\cos (k_n x)$ and $\cos (k_{n-m}x)$, hence  the coefficients $A_n$ and $B_n$ are of order unity, whereas $C_n$ is small, of order $t$.
Then we have, up to a global normalization constant, $A_n \sim \sin \theta_n$ and $B_n \sim \cos \theta_n$, where
\begin{equation}
\tan \theta_n =  2 \frac{E_n - k^2_{n-m}}{t E_n}.
\end{equation}
Away from the degeneracy point, only $A_n$ remains of order unity, whereas $B_n$ decreases and becomes of order $t$, of the same order as the coefficient $C_n$. Therefore, up to linear order in $t$, we have $A_n = D_n \sin \theta_n$, $B_n = D_n  \cos \theta_n$ and $C_n = D_n t E_{n} \sin \theta_n/[2(E_{n}-k_{n+m}^{2})]$,
where $D_n$ is a global normalization constant determined by imposing the weighted normalization condition, see Eq.~(\ref{eq:[c]orthogonality condition Psi}),
\begin{equation}
\int_{0}^{L} [e_c(x)]^{-1} \Psi_{n}^2\left(x\right)  dx=1.\label{eq:[c]norm condition Psi}
\end{equation}
We find, to order $t$, $D_n = \sqrt{2 e_c/L} [1 + t \sin (2 \theta_n)/4]$.

With the same accuracy we obtain $\lambda_n$, using Eq.~(\ref{eq:lambda_n}) and setting $T=0$,
\begin{eqnarray}
\lambda_{n} & = & \sqrt{\frac{\pi}{g\sqrt{L^{2}E_{n}}}}\left(1+\frac{\sin2\theta_{n}}{4}t\right)\nonumber \\
 &  & \times\left[\sin\theta_{n}\left\{ \left(-1\right)^{n}-1\right\} +\cos\theta_{n}\left\{ \left(-1\right)^{n-m}-1\right\} +\frac{tE_{n}/2}{E_{n}-k_{n+m}^{2}}\sin\theta_{n}\left\{ \left(-1\right)^{n+m}-1\right\} \right].\label{eq:[c]lamda general}
\end{eqnarray}
For $t=0$ we recover the result (\ref{eq:lambda_n_hom}) for the homogenous metamaterial. For the modulated wire, the renormalized Josephson energy can be determined numerically using (\ref{eq:[c]lamda general}), summing $\lambda_n^2$, as long as $n_\mathrm{max} \ll m/t$. Interestingly, as can be seen from Eq.~(\ref{eq:[c]lamda general}), the corrections to $\lambda_{n}^{2}$ are of order $t^2$ for odd $m$, whereas they are of order $t$ when $m$ is even. This implies that for even $m$, the renormalization of $E_J$ can be either stronger or weaker than the one found in the homogeneous case, depending on the sign of $t$. As we will show below, this parity effect is general and holds also when $n_\mathrm{max} > m/t$.

\begin{figure}[h]
\includegraphics[scale=0.6,angle=0]{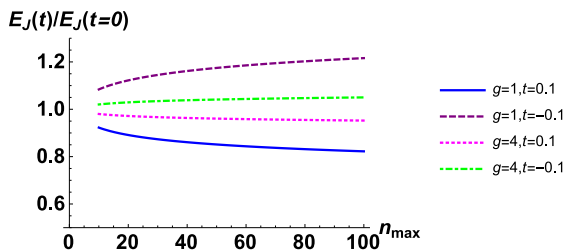}
\caption{(Color online) Ratio $E_{J}(t)/E_{J}(t=0)$, computed from Eq.~(\ref{eq:apprren}), as a function of $n_\mathrm{max}$. Curves from bottom to top correspond to $g=1, t=0.1$ (solid blue),  $g=4, t=0.1$ (dotted pink), $g=4, t=-0.1$ (dashed-dotted green), and $g=1, t=-0.1$ (dashed purple). \label{fig:ejrel}}
\end{figure}

Analytical results can be obtained in the limit $t \ll 1/m$, when the width of the gap region vanishes, $\delta n \ll 1$. We then approximate $E_{n}\simeq k_{n}^{2}$,
$\sin\theta_{n}\simeq1$, and $\cos\theta_{n}\simeq (t n^{2}/2)/[n^{2}-\left(n-m\right)^{2}]$. Taking $m$ even, we find a nonvanishing value $\lambda_n$ for odd values $2l+1$ of $n$ only, hence
\begin{equation}
\lambda_{2l+1}^{2}  = \frac{4}{g\left(2l+1\right)}\left[1+\frac{t\left(2l+1\right)^{2}}{\left(2l+1\right)^{2}-\left(2l+1-m\right)^{2}}
+\frac{t\left(2l+1\right)^{2}}{\left(2l+1\right)^{2}-\left(2l+1+m\right)^{2}}\right].\label{eq:[c]integrand}
\end{equation}
This is valid except for the point $n = 2l+1 = m/2$ where the gap in the spectrum opens up. At the gap, according to Eq.~(\ref{eq:[c]lamda general}),
$\lambda_{m/2}^2 \sim 1/m$, which we neglect assuming $m \gg 1$.
We then evaluate
\begin{equation}
\sum_{l=0,2l+1 \ne m/2}^{l_{\mathrm{max}}}\lambda_{2l+1}^{2}
 =  (2/g)[\log 4 + \gamma + (1 +t/2)\log l_\mathrm{max}], \label{eq:pertlambda2}
\end{equation}
where we dropped contributions of order $1/m$ and $1/l_\mathrm{max} \simeq 2/n_\mathrm{max}$ and assumed $n_\mathrm{max} \gg m$. Result (\ref{eq:pertlambda2}) is obtained using results (\ref{eq:[c]replacementoddonly}), (\ref{eq:[c]replacement2shifted}) and (\ref{eq:[c]replacement4shifted}) from Appendix \ref{sec:sums}, together with the identity
\begin{equation}
\frac{2l+1}{\left(2l+1\right)^{2}-\left(2l+1-m\right)^{2}}
+\frac{2l+1}{\left(2l+1\right)^{2}-\left(2l+1+m\right)^{2}} = \frac{1}{2}\left(\frac{1}{2l+1-m}
+\frac{1}{2l+1+m}\right).
\end{equation}
As a result
\begin{equation}
E_J(t) = E_J^0 (1/4 e^\gamma)^{1/g} (2/n_\mathrm{max})^{(1+t/2)/g}.
\label{eq:apprren}
\end{equation}
For $t=0$ we recover the result for a homogeneous metamaterial. In Fig.~\ref{fig:ejrel} we plot the ratio $E_J(t)/E_J(t=0)$, obtained from Eq.~(\ref{eq:apprren}), of the renormalized Josephson energies found for the modulated and the homogeneous metamaterial as a function of $n_\mathrm{max}$ for $t= \pm 0.1$ and for $g=1$ and $g=4$. Indeed, depending on the sign of $t$, the renormalization of $E_J$ for a modulated metamaterial will be either stronger or weaker as compared to the homogeneous case. Since these results are obtained in the limit $t \ll 1/m$, where the width of the gap region vanishes, $\delta n \ll 1$, we conclude that the effects of the modulation on the renormalization of the Josephson energy are due to the linear in $t$ corrections to the modes, and not due to the opening of the gap in the spectrum. This is confirmed by the fact that the difference between the modulated and homogeneous metamaterials grows with $n_\mathrm{max}$: the larger the number of modes involved in the renormalization, the larger the effect of the modulation.

\subsection{The WKB approximation}

As we have seen above, the perturbation theory with respect to $t$ breaks down for modes $n > m/t$. On the other hand, for modes with large $n$ we can use a quasi-classical procedure to obtain the spectrum and the eigenfunctions of the wave equation (\ref{eq:[c]equation2}). In this subsection, we use the WKB
approximation\cite{Landau} and solve Eq.~(\ref{eq:[c]equation2}) for large energies $E_n$. In this approximation, the mode $\Psi_{n}\left(x\right)$ can be expressed as
\begin{equation}
\Psi_{n}\left(x\right)=\frac{1}{\sqrt{p_{n}\left(x\right)}}\left(F_{n}e^{i\int_{0}^{x}p_{n}\left(x^{\prime}\right)
dx^{\prime}}+G_{n}e^{-i\int_{0}^{x}p_{n}\left(x^{\prime}\right)dx^{\prime}}\right),\label{eq:[c]WKB wave function}
\end{equation}
where $p_n$ is the quasi-classical momentum,
\begin{equation}
p_{n}\left(x\right)=\sqrt{E_{n}\left(1-t\cos k_{m}x\right)}.\label{eq:[c]WKB momentum}
\end{equation}
The coefficients $F_n = G_n$ in order to satisfy $d\Psi_{n}/dx |_0 =0$. Thus
\begin{equation}
\Psi_{n}\left(x\right)=\frac{2}{\sqrt{p_{n}\left(x\right)}}F_{n}\cos \int_{0}^{x} p_{n}\left(x^{\prime}\right)dx^{\prime}.\label{eq:[c]WKB cos_wave function}
\end{equation}
Imposing $d\Psi_{n}/dx |_L =0$, we obtain the quantization condition $\int _0 ^L dx p_{n}\left(x\right) = n \pi$. Thus the eigenvalues $E_n$ satisfy
\begin{equation}
\sqrt{E_{n}}=\frac{n\pi^{2}}{2L\sqrt{1-t}E[2t/(t-1)]} \approx k_{n}[1+\frac{t^{2}}{16}+\mathcal{O}(t^3)],
\label{eq:[c]energy WKB}
\end{equation}
where $E(k)$ is the complete elliptic integral of the second
kind,
\begin{equation}
E\left(k\right)=\int_{0}^{\frac{\pi}{2}}\sqrt{1-k\sin^{2}x}dx\label{eq:elliptic integral 2nd},
\end{equation}
and where we used the asymptotic expression\cite{Abramowitz}
\begin{equation}
E\left(k\right)\approx\frac{\pi}{2}
\left[1-\left(\frac{1}{2}\right)^{2}\frac{k}{1}-\left(\frac{1\cdot3}{2\cdot4}\right)^{2}
\frac{k^{2}}{3}-\left(\frac{1\cdot3\cdot5}{2\cdot4\cdot6}\right)^{2}\frac{k^{3}}{5}+\mathcal{O}\left(k^{4}\right)\right],
\end{equation}
valid for $k\ll 1$, to obtain the approximate result for $\sqrt{E_n}$ valid when $\left|t\right|\ll1$.

Next we use the weighted normalization condition Eq.~(\ref{eq:[c]norm condition Psi}) to obtain the constant $F_n$.
Imposing
\begin{equation}
\frac{4 F_n^2}{e_{c}\sqrt{E_n}}\int_{0}^{L}\cos^{2}\left\{ \sqrt{E_{n}}\int_{0}^{x}\sqrt{1-t\cos k_{m}x^{\prime}}dx^{\prime}\right\} \sqrt{1-t\cos k_{m}x}dx=1,\label{eq:[c]normalization condition WKB}
\end{equation}
defining $s\left(x\right)=\int_{0}^{x}\sqrt{1-t\cos k_{m}x^{\prime}}dx^{\prime}$, and using the fact that
 $s\left(0\right)=0$, $s\left(L\right)=2L\sqrt{1-t}E\left(\frac{2t}{t-1}\right)/\pi$,
we finally obtain
\begin{equation}
\Psi_{n}\left(x\right)=\sqrt{\frac{\pi e_{c}}{L\sqrt{1-t}E\left(\frac{2t}{t-1}\right)}}\frac{\cos \sqrt{E_{n}}s(x)  }{\left(1-t\cos k_{m}x\right)^{\frac{1}{4}}}.\label{eq:[c]WKB normarlized wave function}
\end{equation}
Note that both the energy eigenvalue Eq.~(\ref{eq:[c]energy WKB})
and the wave function Eq.~(\ref{eq:[c]WKB normarlized wave function})
are valid at all orders of $t$.

The validity condition of the WKB approximation used here reads $dp/dx\ll p^{2}$~[\onlinecite{Landau}].
In our case, when $\left|t\right|\ll1$, this condition becomes $n \gg m t$. On the other hand, as can be seen from Eq.~(\ref{eq:[c]energy WKB}), the WKB eigenvalues $E_n$ do not reproduce the gap at $n = m/2$. We thus will use the WKB method only if
\begin{equation}
n\gg m .\label{eq:[c]WKB validity}
\end{equation}
We see that, interestingly, there is a window $m  \ll n \ll m/t$ where both the perturbation theory and the WKB approach hold. As we will see below, this enables us to find the renormalized Josephson energy $E_J$ in a broad range of parameters, combining both approximations.

We finally turn to the computation of $\lambda_n$, Eq.~(\ref{eq:lambda_n}), using the WKB result (\ref{eq:[c]WKB normarlized wave function}). At zero temperature it reads
\begin{equation}
\lambda_{n}=\sqrt{\frac{1}{g n}}\left[\frac{\left(-1\right)^{n}}{\left\{ 1-t\left(-1\right)^{m}\right\} ^{\frac{1}{4}}}-\frac{1}{\left(1-t\right)^{\frac{1}{4}}}\right].\label{eq:[c]WKB lamda}
\end{equation}
For $t=0$, we recover the homogeneous result (\ref{eq:lambda_n_hom}).
When $m$ is an even number, we have
\begin{equation}
\lambda_{n}^2=\frac{2}{g n} \frac{1-\left(-1\right)^{n}}{(1-t)^ {1/2}}. \label{eq:[c]WKB lamda_2_m_even}
\end{equation}
This is of order $t$ for small $t$, hence we expect that the renormalization of the Josephson energy depends on the sign of $t$ in this case.
When $m$ is an odd number, $\lambda_n^2$
starts from order $t^{2}$. This parity effect is the same as the one
found when using perturbation theory with respect to $t$.

\subsection{Renormalized Josephson energy}
\label{sec:renjosen}

We proceed and evaluate $\left\langle \cos[\chi\left(L\right)-\chi\left(0\right)]\right\rangle $. We will combine the results obtained using perturbation theory ($P$) and WKB theory ($W$). We introduce the intermediate point $n^{*}=\sqrt{m \cdot m/t}=m/\sqrt{t}$ within the interval $m < n < m/t$ where both approaches are valid. Then
\begin{eqnarray}
\left\langle \cos[\chi\left(L\right)-\chi\left(0\right)]\right\rangle  & = & e^{-\frac{1}{2}\sum_{n=1}^{n_{\mathrm{max}}}\lambda_{n,W}^{2}}e^{-\frac{1}{2}\sum_{n=1}^{n^{*}}\left(\lambda_{n,P}^{2}-\lambda_{n,W}^{2}\right)}\nonumber \\
 & = & \left\langle \cos[\chi\left(L\right)-\chi\left(0\right)]\right\rangle _{W}\left\langle \cos[\chi\left(L\right)-\chi\left(0\right)]\right\rangle _{P-W}.\label{eq:[c]dividing}
\end{eqnarray}
First we consider $\left\langle \cos[\chi\left(L\right)-\chi\left(0\right)]\right\rangle _{W}=\exp[-(1/2)\sum_{n=1}^{n_{\mathrm{max}}}\lambda_{n,W}^{2}] = (1/2 e^\gamma n_\mathrm{max})^{1/g\sqrt{1-t}}$, where we used (\ref{eq:[c]replacement1}) and (\ref{eq:[c]replacement2}).
Compared to the homogeneous case, we see that $g$ has been replaced by $g\sqrt{1-t}$. Next we consider $\left\langle \cos[\chi\left(L\right)-\chi\left(0\right)]\right\rangle _{P-W}=\exp[-(1/2)\sum_{n=1}^{n^{*}}\left(\lambda_{n,P}^{2}-\lambda_{n,W}^{2}\right)]$, where $\lambda_{n,P}^{2}$ is given by Eq.~(\ref{eq:[c]lamda general}) and $\lambda_{n,W}^{2}$ by Eq.~(\ref{eq:[c]WKB lamda_2_m_even}). In the range where the perturbation theory and the WKB approach are both valid, we use the latter with order $t$ accuracy, hence
\begin{equation}
\lambda_{2l+1,W}^{2} =  \frac{4}{g \left(2l+1\right)}\left(1+\frac{t}{2}\right)\label{eq:[c]lamda W}.
\end{equation}

\begin{figure}[h]
\includegraphics[scale=0.7,angle=0]{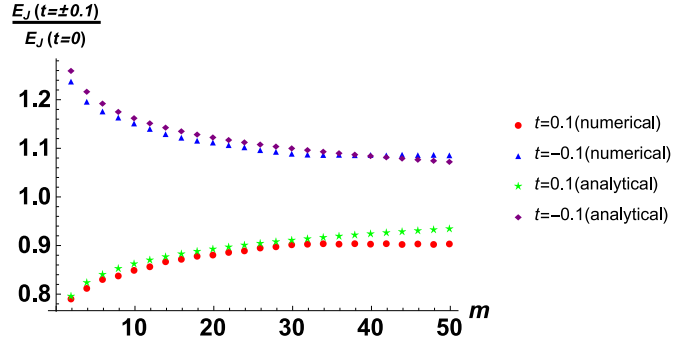}
\caption{(Color online) Ratio $E_{J}(t)/E_{J}(t=0)$ as a function of modulation wavenumber $m$ for a metamaterial with  $n_\mathrm{max} = 100$ and $g=1$ whose capacitance is periodically modulated  with amplitude $t=\pm0.1$. Both numerical results (red dots and blue triangles) and analytical results (green stars and purple diamonds) are shown.\label{fig:renjoscharge}}
\end{figure}
The various sums can be computed numerically, some results are shown in Fig.~\ref{fig:renjoscharge}, where we plot $\ln E_J(t)/E_J(t=0)$ as a function of $m$ for $n_\mathrm{max} = 100$, $g=1$ and $t=\pm 0.1$. First of all, we see that the modulation modifies the result for $E_J$ significantly (by about 20 \%) as compared to the homogeneous case. Also note the dependence on the sign of $t$. The effect is strongest for slow modulation; as $m$ increases, the effect of the modulation weakens. This can be understood as the result of an effective averaging: a modulation with wave number $m$ affects only those modes $\Psi_n$ whose wave numbers $n$ are larger than $m$. For modes with wave numbers $n < m$, the effect of the modulation of the capacitance along the metamaterial just averages out.

Analytical results can be obtained in the limit $t\ll 1/m$. Calculating $\lambda_{n,P}^{2}$ using Eq.~(\ref{eq:[c]integrand}),
we find
\begin{equation}
\lambda_{2l+1,P}^{2}-\lambda_{2l+1,W}^{2}=(2t/g)\left\{1/[2\left(2l+1\right)-m]+1/[2\left(2l+1\right)+m]-1/(2l+1)\right\}.
\label{eq:difflambdaPW}
\end{equation}
Using (\ref{eq:[c]replacementoddonly}), (\ref{eq:[c]replacement2shifted}) and (\ref{eq:[c]replacement4shifted}),
we obtain
\begin{equation}
e^{-\frac{1}{2}\sum_{n=1}^{n^{*}}\left(\lambda_{n,P}^{2}-\lambda_{n,W}^{2}\right)}=\left(me^{\gamma}\right)^{\frac{t}{2g}},\label{eq:[c]difference P-W}
\end{equation}
where we dropped again contributions of order $1/m$ and $1/l_\mathrm{max} \simeq 2/n_\mathrm{max}$ and assumed $n_\mathrm{max} \gg m$.
As a result,
\begin{equation}
\frac{E_{J}(t)}{E_{J}(t=0)}=\left(\frac{m}{2n_{\mathrm{max}}}\right)^{\frac{t}{2g}}\label{eq:[c]cosine factor P+W at order t}
\end{equation}
This result is also shown in Fig.~\ref{fig:renjoscharge} for $g=1$ and $t = \pm 0.1$ (green stars and purple diamonds). We see that the approximate analytical result is fairly accurate for values of $m \alt 30$. This is more or less expected, as the condition $t < 1/m$ breaks down when $m \agt 10$. However, the deviations between analytical and numerical results remain relatively small over the plotted range. This confirms the fact that the effect of the modulation is mainly due to the renormalization of the mode wave functions and not due to the opening of the gap in the mode spectrum.

\section{Discussion}

\begin{figure}[h]
\includegraphics[scale=0.7,angle=0]{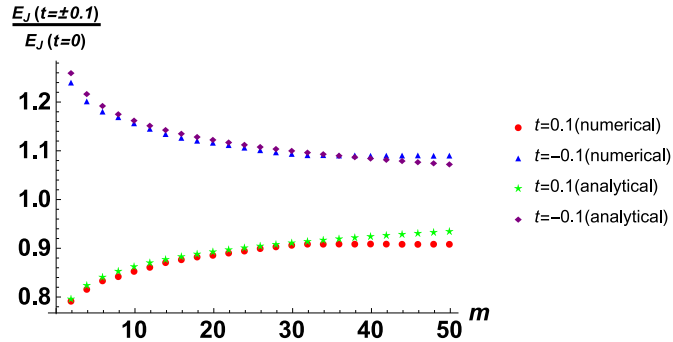}
\caption{(Color online) Ratio $E_{J}(t)/E_{J}(t=0)$ as a function of modulation wavenumber $m$ for a metamaterial with  $n_\mathrm{max} = 100$ and $g=1$ whose inverse inductance is periodically modulated with amplitude $t=\pm0.1$. Both numerical results (red dots and blue triangles) and analytical results (green stars and purple diamonds) are shown.\label{fig:renjosind}}
\end{figure}

We verified that the results found above are not specific for a capacitance modulation. Indeed, introducing a modulation of the inductance $l$ rather than of the capacitance $c$ does not alter our conclusions. In Fig.~\ref{fig:renjosind} we show the results for the renormalization of the Josephson energy $E_J$ for a metamaterial with $c=c_0$ constant and $1/l(x) = (1/l_0)(1-\cos k_m x)$. Details of the calculations are presented in Appendix~\ref{sec:Modulating-inductance}. The results are very similar to the ones presented in Fig.~\ref{fig:renjoscharge}, both qualitatively and quantitatively. Specifically, we find that the approximate analytical result for this case coincides exactly with the one found for the capacitance modulation, Eq.~(\ref{eq:[c]cosine factor P+W at order t}). The numerical result does not deviate strongly from the analytical result as long as $m \alt 30$.

\begin{figure}[h]
\includegraphics[scale=0.7,angle=0]{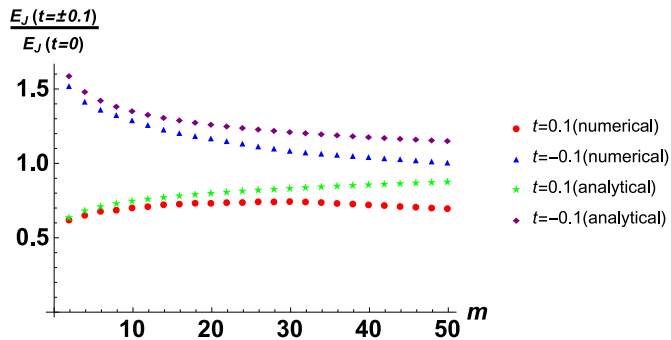}
\caption{(Color online) Ratio $E_{J}(t)/E_{J}(t=0)$ as a function of modulation wavenumber $m$ for a metamaterial with  $n_\mathrm{max} = 100$ and $g=1$ whose capacitance and inverse inductance are modulated simultaneously with amplitude $t=\pm0.1$. Both numerical results (red dots and blue triangles) and analytical results (green stars and purple diamonds) are shown.\label{fig:renjosboth}}
\end{figure}

We also investigated the case when the capacitance and the inductance are modulated simultaneously, such that $c(x) = c_0(1 - t \cos k_m x)$ and $1/l(x) = (1/l_0)(1 - t \cos k_m x)$. Such a modulation leaves the plasma velocity constant, $v_\mathrm{pl}(x) = v_\mathrm{pl} = \sqrt{1/l_0 c_0}$. It corresponds, {\em e.g.}, to a periodic modulation of the lateral size of the islands of a Josephson junction chain, which affects the capacitance to ground $C_g$ and the Josephson coupling $E_{J,ch}$ to the neighbouring islands in the same way, the latter being proportional to $1/L_{ch}$.  Details of the calculations are presented in Appendix~\ref{sec:Modulating both}, the results are shown in Fig.~\ref{fig:renjosboth}. The results are qualitatively similar to the ones presented in Figs.~\ref{fig:renjoscharge} and \ref{fig:renjosind}. Note however the quantitative difference: on a logarithmic scale, the effect of the modulation is twice stronger when both $c$ and $l$ are modulated. This is because both modulations contribute equally to the renormalization of the eigenfunctions $\Psi_n$. Indeed, comparing the approximate analytical result (\ref{eq:[c+l]cosine factor P+W at order t}) with Eq.~(\ref{eq:[c]cosine factor P+W at order t}), we see that the modulation amplitude $t$ is multiplied by a factor 2. The fact of having two contributions to the modulation also causes the difference between the approximate analytical result and the numerical result to grow faster with increasing $m$ as compared to Figs.~\ref{fig:renjoscharge} and \ref{fig:renjosind}.

In order to see whether the effects discussed here are experimentally accessible, we give some typical values for the relevant parameters characterizing currently available metamaterials. Superconducting nanowires~\cite{Astafiev12,Peltonen13} are characterized by an inductance per unit length of about 0.002 H/m. The capacitance per unit length is about that of the vacuum, $c \simeq 100$pF/m, but it can be strongly enhanced when putting the wire on dedicated substrates. For instance, using $SrTi0_3$~\cite{Camarota01} with $\epsilon = 10^4$ as a substrate, one can achieve $c \simeq 10^6$pF/m. We thus have a plasma velocity in the range $v_\mathrm{pl} \simeq 10^4 $ -- $10^6$m/s and a dimensionless conductance $g \simeq 1$ -- 100. Assuming the gap to be in the 10 Kelvin range, such wires sustain about $10^6$ -- $10^8$ modes per meter. For a wire length of about 10$\mu$m, we have 10 - 1000 modes. With these values, a 10 \% spatial modulation of the nanowire parameters on a $\mu$m scale would realistically yield results similar to those shown in Figs.~\ref{fig:renjoscharge}, \ref{fig:renjosind}, and \ref{fig:renjosboth}. Josephson junction chains ~\cite{Ergul13,Weissl15} may well have lengths $N$ up to several thousands of junctions. Due to the relatively small capacitance to ground (typically, the ratio $C_g/C_{ch} \simeq 0.01$), the characteristic plasma velocities are relatively high, such that the number of available propagating plasma modes $n_\mathrm{max} \sim  N \sqrt{C_g/C_{ch}}$ typically is only about 10 \% of the chain's length. For the same reason, $g \simeq 1$ is relatively small. Nevertheless, with these numbers a 10 \% spatial modulation of the chain's parameters over a few junctions would again realistically yield results similar to the ones obtained above.

In conclusion, we studied the renormalization of the Josephson energy of a small Josephson junction, embedded in a loop made out of a one-dimensional metamaterial. We found that a slow modulation of the metamaterial's parameters such as its capacitance or its inductance per unit length has a strong effect on the renormalized Josephson energy. Depending on the sign of the modulation amplitude, the renormalization can be either significantly stronger or significantly weaker than the one obtained for a homogeneous superconducting loop~\cite{Hekking97}. The modulation affects the electromagnetic modes propagating along the metamaterial as well as their frequency spectrum. However, interestingly, the effect of the modulation on the renormalization of the Josephson energy is mainly due to the modified behaviour of the mode spatial structure close to the Josephson junction, than due to the modified frequency spectrum. This example illustrates how mode engineering with a superconducting metamaterial can be used to affect its coupling to another superconducting device of interest in a controllable way.

\acknowledgments
The authors thank O. Buisson, W. Guichard, S. Kawabata Y. Krupko, S. Nakamura, N. Roch, and
Y. Tokura for discussions. M.T. thanks LPMMC and Institut N$\acute{\mathrm{e}}$el for hospitality and acknowledges
support from the Tsukuba Nanotechnology Human Resource Development Program. We also acknowledge financial support from Institut universitaire de France and the European Research council (grant no. 306731).

\appendix

\section{Some useful sums}
\label{sec:sums}
In this Appendix we present results for the sums encountered in the main text. We start from the elementary sum\cite{Gradshteyn}
\begin{equation}
\sum_{n=1}^{n_{\mathrm{max}}}\frac{1}{n}  =  \ln n_{\mathrm{max}}+\gamma+\mathcal{O}(1/n_{\mathrm{max}}),\label{eq:[c]replacement1}
\end{equation}
where $\gamma = 0.5772 \ldots$ is Euler's constant. When summing over odd values of $n$ only, we have\cite{Gradshteyn}
\begin{equation}
\sum_{l=0}^{l_{\mathrm{max}}}\frac{1}{2l+1}  =  (1/2) \ln l_{\mathrm{max}}+\gamma/2+ \ln 2 + \mathcal{O}(1/l_{\mathrm{max}}).\label{eq:[c]replacementoddonly}
\end{equation}
From (\ref{eq:[c]replacement1}) we see that a sum involving only the even values of $n$ yields
\begin{equation}
\sum_{l=1}^{l_{\mathrm{max}}}\frac{1}{2l}  =  (1/2) \ln l_{\mathrm{max}}+\gamma/2+ \mathcal{O}(1/l_{\mathrm{max}}).\label{eq:[c]replacementevenonly}
\end{equation}
Using (\ref{eq:[c]replacementoddonly}) and (\ref{eq:[c]replacementevenonly}), we obtain
\begin{equation}
\sum_{n=1}^{n_{\mathrm{max}}}\frac{\left(-1\right)^{n+1}}{n}  =  \ln2+\mathcal{O}(1/n_{\mathrm{max}}).\label{eq:[c]replacement2}
\end{equation}

We next consider the sum for integer $m>0$
\begin{equation}
\sum \limits _{n=1, n \ne m}^{n_{\mathrm{max}}} \frac{1}{n -m}  = - \sum \limits _{p=1}^{m-1} \frac{1}{p} + \sum \limits _{p=1}^{n_{\mathrm{max}}-m} \frac{1}{p} = \ln (n_{\mathrm{max}}-m) - \ln m+\mathcal{O}[1/(n_{\mathrm{max}}-m),1/m],
\label{eq:[c]replacement1shifted}
\end{equation}
where we used (\ref{eq:[c]replacement1}). We use this result to compute
\begin{equation}
\sum \limits _{l=1, 2l-1 \ne m/2}^{l_{\mathrm{max}}} \frac{1}{2(2l-1) - m}  = (1/4)[\ln (n_{\mathrm{max}}-m/4) - \ln (m/4)] +\mathcal{O}[1/(n_{\mathrm{max}}-m/4),1/m].
\label{eq:[c]replacement2shifted}
\end{equation}
Similarly, for integer $m>0$,
\begin{equation}
\sum \limits _{n=1}^{n_{\mathrm{max}}} \frac{1}{n + m}  = - \sum \limits _{p=1}^{m} \frac{1}{p} + \sum \limits _{p=1}^{n_{\mathrm{max}}+m} \frac{1}{p} = \ln (n_{\mathrm{max}}+m) - \ln m+\mathcal{O}[1/(n_{\mathrm{max}}+m),1/m],
\label{eq:[c]replacement3shifted}
\end{equation}
such that
\begin{equation}
\sum \limits _{l=1}^{l_{\mathrm{max}}} \frac{1}{2(2l-1) + m}  = (1/4)[\ln (n_{\mathrm{max}}+m/4) - \ln (m/4)] +\mathcal{O}[1/(n_{\mathrm{max}}+m/4),1/m].
\label{eq:[c]replacement4shifted}
\end{equation}

\section{Modulating the kinetic inductance \label{sec:Modulating-inductance}}

In this Appendix, we consider the case when the kinetic inductance $l(x)$ is space-dependent and the capacitance $c=c_0$ is constant.
Equation~(\ref{eq:Psi_n}) then reduces to
\begin{equation}
\frac{d}{dx}\left[v_\mathrm{pl}^2\left(x\right)\frac{d \Psi_n(x)}{dx} \right] + \omega_{n}^{2}\Psi_{n}\left(x\right) = 0,
\label{eq:[s]equation1}
\end{equation}
where $v_\mathrm{pl}^2(x) = 1/c_0 l(x)$.
Now we consider a periodic modulation for the inductance such that $1/l(x) = (1/l_0)(1 - t \cos k_m x)$, in analogy with the capacitance modulation (\ref{eq:modulating capacitance}). As a result,  $v_\mathrm{pl}^2(x) = v_\mathrm{pl}^2 (1 - t \cos k_m x)$, where $v_\mathrm{pl}^2 = 1/l_0c_0$. We then can rewrite Eq.~(\ref{eq:[s]equation1}), such that
\begin{equation}
\left(1-t\cos k_{m}x\right)\frac{d^{2}\Psi_{n}\left(x\right)}{dx^{2}}+tk_{m}\sin k_{m}x\frac{d\Psi_{n}\left(x\right)}{dx}+E_{n}\Psi_{n}\left(x\right)=0,\label{eq:[s]equation2}
\end{equation}
where we defined $E_{n}=(\omega_n/v_\mathrm{pl})^2$.

We first use perturbation theory ($P$) and repeat the steps outlined in Sec.~\ref{subsec:perturbation} for the capacitance modulation. The modes $\Psi_n(x)$ satisfy expansion (\ref{eq:[c]three cosine}), where the coefficients $A_n$, $B_n$ and $C_n$ are now found from the set of equations
\begin{eqnarray}
\left(E_{n}-k_{n}^{2}\right)A_{n}+(t/2)k_{n}k_{n-m}B_{n}+(t/2)k_{n}k_{n+m}C_{n} & = & 0,\label{eq:[s]condition1}\\
\left(E_{n}-k_{n-m}^{2}\right)B_{n}+(t/2)k_{n}k_{n-m}A_{n} & = & 0,\label{eq:[s]condition2}\\
\left(E_{n}-k_{n+m}^{2}\right)C_{n}+(t/2)k_{n}k_{n+m}A_{n} & = & 0,\label{eq:[s]condition3}
\end{eqnarray}
obtained using (\ref{eq:[s]equation2}).
From this we find that the eigenvalues $E_n$ are given again by Eq.~(\ref{eq:eigenvalues}); for the coefficients we find $A_n = \sqrt{2 e_c/L} \sin \theta_n$, $B_n = \sqrt{2 e_c/L} \cos \theta_n$ and $C_n = -t \sqrt{2 e_c/L} \sin \theta_n  k_n k_{n+m}/[2(E_n - k^2_{n+m})]$.
Here, the angle $\theta_n$ satisfies
\begin{equation}
\tan \theta_n = 2 \frac{k^2_{n-m} -E_n}{t k_n k_{n-m}}.
\end{equation}
As a result, the coefficients $\lambda_{n,P}$ read
\begin{equation}
\lambda_{n,P}  =  \sqrt{\frac{\pi}{g\sqrt{L^{2}E_{n}}}}\left[\sin\theta_{n}\left\{ \left(-1\right)^{n}-1\right\}
+\cos\theta_{n}\left\{ \left(-1\right)^{n-m}-1\right\} -\frac{t k_{n}k_{n+m}/2}{E_{n}-k_{n+m}^{2}}\sin\theta_{n}\left\{ \left(-1\right)^{n+m}-1\right\} \right].
 \label{eq:[s]lamda general}
\end{equation}
When $m$ is an odd number, $\lambda_{n,P}^{2}$ is even function with respect to $t$, whereas for even $m$ the correction is of order $t$. In the limit $t \ll 1/m$, we can set $\sin\theta_{n}\simeq 1$ and
$\cos\theta_{n}\simeq-(t/2)n(n-m)/[n^{2}-(n-m)^{2}]$. Then
we obtain
\begin{equation}
\lambda_{2l+1,P}^{2}  =  \frac{4}{g\left(2l+1\right)}\left[1-\frac{t\left(2l+1\right)\left(2l+1-m\right)}{\left(2l+1\right)^{2}-\left(2l+1-m\right)^{2}}
-\frac{t\left(2l+1\right)\left(2l+1+m\right)}{\left(2l+1\right)^{2}-\left(2l+1+m\right)^{2}}\right].\label{eq:[s]integrand}
\end{equation}

The validity condition of the perturbation theory is again $n \ll m/t$. For larger values of $n$ we can use the WKB method ($W$) to obtain the modes $\Psi_n(x)$. In this approximation, the modes are given by Eq.~(\ref{eq:[c]WKB wave function}), where the quasi-classical momentum is now given by
\begin{equation}
p_{n}\left(x\right)=\sqrt{\frac{E_{n}}{1 - t \cos k_m x}}.\label{eq:[s]WKB momentum}
\end{equation}
Imposing the boundary conditions $d\Psi_n/dx|_{0,L}=0$, we see that $F_n = G_n$ and find the quantization condition
\begin{equation}
\sqrt{E_{n}^{s}}=\frac{n\pi^{2}\sqrt{1-t}}{2LK[2t/(t-1)]} \approx k_{n}[1-\frac{3t^{2}}{16}+\mathcal{O}\left(t^{3}\right)],\label{eq:[s]energy WKB}
\end{equation}
where we defined the complete elliptic integral of the first kind
\begin{equation}
K\left(k\right)=\int_{0}^{\frac{\pi}{2}}\frac{1}{\sqrt{1-k\sin^{2}x}}dx,\label{eq:elliptic integral 1st}
\end{equation}
and used the asymptotic expression\cite{Abramowitz} for small values of the argument,
\begin{equation}
K\left(k\right) \approx \frac{\pi}{2}
\left[1+\left(\frac{1}{2}\right)^{2}k+\left(\frac{1\cdot3}{2\cdot4}\right)^{2}k^{2}+
\left(\frac{1\cdot3\cdot5}{2\cdot4\cdot6}\right)^{2}k^{3}+\mathcal{O}\left(k^{4}\right)\right],
\end{equation}
to obtain an approximated value for $E_n$ when $\left|t\right| \ll 1$. Defining,
$r\left(x\right)=\int_{0}^{x}dx'/\sqrt{1-t\cos k_{m}x^{\prime}}$
we finally obtain
\begin{equation}
\Psi_{n}\left(x\right)=\sqrt{\frac{\pi e_{c}\sqrt{1-t}}{L K[2t/(t-1)]}}\frac{\cos \sqrt{E_{n}}r(x)  }{\left(1-t\cos k_{m}x\right)^{\frac{1}{4}}},\label{eq:[s]WKB normarlized wave function}
\end{equation}
imposing the weighted normalization condition (\ref{eq:[c]norm condition Psi}).
For even $m$ and to order $t$, the parameter $\lambda^2_n$ is nonvanishing for odd $n$ only and given by
\begin{equation}
\lambda_{2l+1,W}^{2}  =  \frac{4}{g\left(2l+1\right)}\left(1+\frac{t}{2}\right), \label{eq:[s]lamda W}
\end{equation}
which coincides with the result (\ref{eq:[c]lamda W}) found for the case of a modulated capacitance.

The WKB approximation can be used for large $n \gg t m $. However, as in Sec.~\ref{sec:Modulating capacitance}, we use it for $n \gg m$ in combination with perturbation theory to compute the renormalized Josephson energy. Introducing again $n^* = m/\sqrt{t}$, we compute $\langle \cos[\chi(L)-\chi(0)]\rangle$, Eq.~(\ref{eq:[c]dividing}), numerically with the help of Eqs.~(\ref{eq:[s]lamda general}) and (\ref{eq:[s]lamda W}). An analytical evaluation is possible in the limit $t \ll 1/m$, with the help of (\ref{eq:[s]integrand}). In this limit,
\begin{equation}
\lambda_{2l+1,P}^{2} - \lambda_{2l+1,W}^{2}  =  \frac{2t}{g}\left[\frac{1}{2(2l+1) -m} + \frac{1}{2(2l+1) +m} \right ],
\end{equation}
which coincides with the result (\ref{eq:difflambdaPW}) obtained in Sec.~\ref{sec:renjosen}. Hence the ratio $E_J(t)/E_J(t=0)$ is again given by Eq.~(\ref{eq:[c]cosine factor P+W at order t}).

\section{Modulating both the capacitance and the kinetic inductance\label{sec:Modulating both}}

In this Appendix, we consider a simultaneous modulation of the capacitance and the inductance, such that $c(x) = c_0(1 - t \cos k_m x)$ and $1/l(x) = (1/l_0)(1 - t \cos k_m x)$. Then, Eq.~(\ref{eq:Psi_n}) reads
\begin{equation}
\left(1-t\cos k_{m}x\right)\frac{d^{2}\Psi_{n}\left(x\right)}{dx^{2}}+tk_{m}\sin k_{m}x\frac{d\Psi_{n}\left(x\right)}{dx}+E_{n}\left(1-t\cos k_{m}x\right)\Psi_{n}\left(x\right)=0,\label{eq:[c,s]equation2}
\end{equation}
where we defined $E_{n}= (\omega_n/ v_\mathrm{pl})^2$, with $v^2_\mathrm{pl} = 1/l_0 c_0$.

Using perturbation theory ($P$) as in Sec.~\ref{subsec:perturbation}, we start from the expansion (\ref{eq:[c]three cosine}). Using Eq.~(\ref{eq:[c,s]equation2}), this yields a set of equations for the coefficients $A_n$, $B_n$, and $C_n$,
\begin{eqnarray}
\left(E_{n}-k_{n}^{2}\right)A_{n}+\frac{t}{2}\left(k_{n}k_{n-m}-E_{n}\right)B_{n}+\frac{t}{2}\left(k_{n}k_{n+m}-E_{n}\right)C_{n} & = & 0,\label{eq:[c,s]condition1}\\
\left(E_{n}-k_{n-m}^{2}\right)B_{n}+\frac{t}{2}\left(k_{n}k_{n-m}-E_{n}\right)A_{n} & = & 0,\label{eq:[c,s]condition2}\\
\left(E_{n}-k_{n+m}^{2}\right)C_{n}+\frac{t}{2}\left(k_{n}k_{n+m}-E_{n}\right)A_{n} & = & 0.\label{eq:[c,s]condition3}
\end{eqnarray}
Up to corrections of order $t^2$, the eigenvalues are given by
\begin{equation}
E_{n,\pm} = \frac{(k_n^2 + k_{n-m}^2) \pm \sqrt{(k_n^2 -k_{n-m}^2)^2 + t^2 (2 k_n^2 k_{n-m}^2- k_n^3 k_{n-m} - k_n k^3_{n-m})}}{2}.
\label{eq:eigenvaluessim}
\end{equation}
This yields the gap $2 |t| k^2_{m/2}$ for $n=m/2$.  Up to linear order in $t$, we have $A_n = \sqrt{2 e_c/L} [1 + t \sin (2 \theta_n)/4] \sin \theta_n$, $B_n = \sqrt{2 e_c/L} [1 + t \sin (2 \theta_n)/4] \cos \theta_n$ and $C_n = \sqrt{2 e_c/L} t (E_{n}-k_n k_{n+m}) \sin \theta_n/[2(E_{n}-k_{n+m}^{2})]$,
where
\begin{equation}
\tan \theta_n =  2 \frac{E_n - k^2_{n-m}}{t(E_n -k_n k_{n-m})}.
\end{equation}
As a result,
\begin{eqnarray}
\lambda_{n,P}  =  \sqrt{\frac{\pi}{g\sqrt{L^{2}E_{n}}}}[1 + t \sin (2 \theta_n)/4]  \left[\sin \theta_n\left\{ \left(-1\right)^{n}-1\right\} + \cos \theta_n\left\{ \left(-1\right)^{n-m}-1\right\} \right. \\ \left.+ \sin \theta_n (t/2)\frac{E_n-k_n k_{n+m}}{E_n-k_{n+m}^2}\left\{ \left(-1\right)^{n+m}-1\right\} \right].\label{eq:[c,s]lamda general}
\end{eqnarray}
For even values of $m$ and in the limit $t \ll 1/m$ this yields,
\begin{equation}
\lambda_{2l+1,P}^{2}=\frac{4}{g\left(2l+1\right)}\left[1+t\frac{\left(2l+1\right)}{2\left(2l+1\right)-m}
+t\frac{\left(2l+1\right)}{2\left(2l+1\right)+m}\right].\label{eq:[c,s]lamda}
\end{equation}
The coefficients are nonzero for odd values of $n = 2l+1$ only.

When $n \gg m/t$, perturbation theory breaks down, and we resort again to the WKB approach. We obtain the eigenfunctions
\begin{equation}
\Psi_{n}\left(x\right)=\sqrt{\frac{2e_{c}}{L}}\frac{\cos\left\{ \sqrt{E_{n}}x\right\} }{\sqrt{1-t\cos k_{m}x}},\label{eq:[c,s]normalized wave function WKB}
\end{equation}
which satisfy the weighted normalization condition (\ref{eq:[c]norm condition Psi}), as well as the boundary conditions $d\Psi_n/dx|_{0,L}=0$, provided
\begin{equation}
\sqrt{E_{n}}=\frac{n\pi}{L}.\label{eq:[c,s]WKB energy}
\end{equation}
The coefficients $\lambda_{n,W}$ then read
\begin{equation}
\lambda_{n,W}=\sqrt{\frac{1}{gn}}\left[\frac{\left(-1\right)^{n}}{\sqrt{1-t\left(-1\right)^{m}}}-\frac{1}{\sqrt{1-t}}\right].\label{eq:[c,s]lamda WKB}
\end{equation}
For odd $n = 2l+1$ and to order $t$ we have
\begin{equation}
\lambda_{2l+1,W}^2=\frac{4}{g(2l+1)}(1+t).
\end{equation}
Note that, as compared to Eqs.~(\ref{eq:[c]lamda W}) and (\ref{eq:[s]lamda W}), the amplitude $t$ appears with a factor of 2 here. This reflects the fact that the capacitance and inductance modulation contribute equally to the renormalization of the wave functions $\Psi_n$.

As in Sec.~\ref{sec:renjosen}, we use the WKB approach for $n \gg m$ in combination with perturbation theory to compute the renormalized Josephson energy. Introducing again $n^* = m/\sqrt{t}$, we compute $\langle \cos[\chi(L)-\chi(0)]\rangle$, Eq.~(\ref{eq:[c]dividing}), numerically, now with the help of Eqs.~(\ref{eq:[c,s]lamda general}) and (\ref{eq:[c,s]lamda WKB}). An analytical evaluation is possible in the limit $t \ll 1/m$. In this limit, we use (\ref{eq:[c,s]lamda}) and obtain
\begin{equation}
\lambda_{2l+1,P}^{2} - \lambda_{2l+1,W}^{2}  =  \frac{4t}{g}\left[\frac{1}{2(2l+1) -m} + \frac{1}{2(2l+1) +m} \right ].
\end{equation}
Note that this is twice the result (\ref{eq:difflambdaPW}) found in Sec.~\ref{sec:renjosen}, reflecting again the equal contributions of the capacitance and inductance modulation. The ratio $E_J(t)/E_J(t=0)$ is then given by
\begin{equation}
\frac{E_{J}(t)}{E_{J}(t=0)}=\left(\frac{m}{2n_{\mathrm{max}}}\right)^{\frac{t}{g}},\label{eq:[c+l]cosine factor P+W at order t}
\end{equation}
where the exponent has doubled as compared to Eq.~(\ref{eq:[c]cosine factor P+W at order t}).

\end{document}